\documentclass[letterpaper, 12pt]{article}
\usepackage{latexsym}
\usepackage{amsmath}
\usepackage{amssymb}
\usepackage[latin1]{inputenc}
\usepackage{graphicx}
\newcommand{\be}{\begin{equation}}
\newcommand{\ee}{\end{equation}}
\newcommand{\bea}{\begin{eqnarray}}
\newcommand{\eea}{\end{eqnarray}}

\newcommand{\labell}[1]{\label{#1}\qquad_{#1}} 
\newcommand{\bbibitem}[1]{\bibitem{#1}\marginpar{#1}}
\newcommand{\llabel}[1]{\label{#1}\marginpar{#1}}

\def\Label#1{\label{#1}%
  \smash{\hbox to0pt{\raise1ex\hbox{\tiny[#1]}\hss}}}
\def\noLabels{\let\Label=\label}
\def\nolabells{\let\labell=\label}
\def\nollabels{\let\llabel=\label}
\def\nobbibitem{\let\bbibitem=\bibitem}

\title{{\bf Experimental design and model selection: } \\
{\bf The example of exoplanet detection}}

\author{Vijay Balasubramanian${}^{1,2}$, Klaus
Larjo${}^1$ and Ravi Sheth${}^1$\footnote{vijay@physics.upenn.edu, klarjo@physics.upenn.edu, shethrk@physics.upenn.edu}
\\[1mm]
\small \sl ${}^1$David Rittenhouse Laboratories, University of Pennsylvania,
\\[-1.5mm]
\small \sl Philadelphia, PA 19104, USA
\\[1mm]
\small \sl ${}^2$School of Natural Sciences, Institute for Advanced Study,
\\[-1.5mm]
\small \sl Princeton, NJ 08540, USA \\
}
\date{}

\begin{document}

\noLabels \setlength{\baselineskip}{16pt} \nobbibitem \nollabels

\begin{titlepage}
%
\maketitle

\begin{abstract}
We apply the Minimum Description Length model selection approach to the detection of extra-solar planets, and use this example to
show how specification of the experimental design affects the prior distribution on the model parameter space and hence the
posterior likelihood which, in turn, determines which model is regarded as most `correct'.    Our analysis shows how conditioning
on the experimental design can render a non-compact parameter space effectively compact, so that the MDL model selection problem
becomes well-defined.
\end{abstract}
\thispagestyle{empty} \setcounter{page}{0}
\end{titlepage}

\section{Introduction}
\label{intro}

The Bayesian approach to parametric model selection requires the specification of  a prior probability distribution over the
parameter space.   The Jeffreys' prior, which is proportional to the square root of the determinant of the Fisher information
computed in the parameter space, has been shown to be the uniform prior over all {\it distributions} indexed by the parameters in
a parametric family \cite{vijay}.   Geometrically, its integral over a region of the parameter space computes a volume that
essentially measures the fraction of statistically distinguishable probability distributions within that region \cite{vijay}. In
this interpretation, the Jeffreys prior distribution
\begin{equation}
\omega(\Theta) = \frac{\sqrt{\textrm{det }J_{ij}(\Theta)}}{\int d^d\Theta \sqrt{\textrm{det }J_{ij}(\Theta)}} d^d\Theta
\end{equation}
where $\Theta = \{\theta_1, \cdots, \theta_d \}$ simply measures the fractional volume of the small element $d^d\Theta$ relative
to total volume of the parametric manifold $V = \int d^d\Theta \, \sqrt{\det J_{ij}(\Theta)}$.   Here $J_{ij}$ is the Fisher
information on the parameter space $\Theta \in \mathbb{R}^d$ and $d^d\Theta$ is the standard Riemannian volume element on
$\mathbb{R}^d$.   The volume $V$ also appears in the Minimum Description Length (MDL) approach to model selection
\cite{rissanen78,rissanen96}, conceptually because it effectively measures how many different distributions are describable by
different parameter choices.

An important difficulty in applying the MDL approach to model selection occurs when the parameter space is noncompact and the
volume $V$ diverges.  In this case, from the Bayesian perspective, a uniform prior on the parameter space does not exist, while
from the MDL perspective the number of models that might be describable diverges, leading to problems with the definition of the
description length.    Of course the parameter space can be cut off by hand, but unless the choice of cut-off is well founded, it
can lead to artifacts in the comparison of different model families \cite{myungetal,liangbarron,grunwald}.   Unfortunately in
many practical problems the parameter space {\it is} noncompact and $V$ diverges.   For example, in astrophysics, the detection of
exoplanets depends on a model of the light coming from the occluded star.  This model will contain a non-compact direction
representing the orbital period of the planet -- see, e.g., \cite{raul}.  For examples from psychophysics see, e.g.,
\cite{myungetal}.

In this note we argue that merely specifying the experimental set-up -- before the measurement of any actual data -- influences
the prior distribution on the parameter space.     This occurs because, given the finite number of measurements in any
experiment, many of the probability distributions indexed by a parametric manifold will be statistically indistinguishable.  In
cases where the parameter space is noncompact, the uniform prior conditioned on the experimental setup can thus become
well-defined.  In the geometric language of \cite{vijay}, the volume that measures the number of probability distributions in the
parametric family that are statistically distinguishable given a {\it finite} number of measurements can be finite even if the
parameter space is non-compact.   In effect, specifying the experimental set-up can render the parameter space compact.

Our results  illustrate how  the choice of experimental set-up influences the measure on the parameter space of a model,
thereby affecting which model is regarded as most `correct'.   In section \ref{general-argument}  we briefly review the
computation of posterior probabilities, and consider the effect of conditioning on the experimental set-up  on the parameter space measure. In section
\ref{sec:example} we apply these considerations to a physical problem: the analysis of light-curves of stars with orbiting
planets. In this example we see that the volume of the parameter space is rendered effectively finite after the experimental
set-up is specified.

\section{The effect of experimental design on the parameter space measure}
\llabel{general-argument}
\subsection{Review}
Suppose one is interested in some physical phenomenon, and has made $N$ relevant measurements: $Y = \{y_1, \ldots, y_N\}$.
Further suppose that there are two different parametric models, $A$ and $B$, that aim to describe the phenomenon in question. The
basic question to be answered is which of the two models is the better one, considering the experimental data $Y$. The
probability-theoretic answer to this question is to compute the posterior probabilities $P(A|Y)$ and $P(B|Y)$, which we can write
using the Bayes Rule as
\begin{equation}
P(A|Y) = \frac{P(A)}{P(Y)} \int \omega(\Theta) P(Y|\Theta), \Label{bayes}
\end{equation}
where $\Theta = (\theta_1,\ldots,\theta_d) \in \mathbb{R}^d$ is the vector of variables parametrising $A$, and $\omega(\Theta)$ is the volume form
associated to the measure on the parameter space, which we will define shortly. A corresponding expression can also be written
for $P(B|Y)$. Since we wish to compare $P(A|Y)$ and $P(B|Y)$, we can ignore the common factor $P(Y)$, and we will assume $P(A) =
P(B)$ and drop this factor as well. Thus the only remaining ingredient to be defined is the volume form $\omega(\Theta)$; we
simply quote the result from \cite{vijay}: the volume form that gives equal weight to all statistically distinguishable
distributions in the parametric family is
\begin{equation}
\omega(\Theta) = \frac{\sqrt{\textrm{det }J_{ij}(\Theta)}}{\int d^d\Theta \sqrt{\textrm{det }J_{ij}(\Theta)}} d^d\Theta,
\Label{measure}
\end{equation}
where $J_{ij}(\Theta)$ is the {\it Fisher information matrix}, defined as the second derivative of the Kullback--Leibler distance
$D(\Theta_p||\Theta_q)$:
\begin{eqnarray}
J_{ij}(\Theta_p) &  = & \partial_{\theta_i} \partial_{\theta_j} D(\Theta_p||\Theta_p + \Phi)|_{\Phi=0}, \Label{fisher-orig}\\
D(\Theta_p || \Theta_q ) & = & \int d\vec{x}\,\,  \Theta_p(\vec{x}) \ln \frac{\Theta_p(\vec{x})}{\Theta_q(\vec{x})}. \Label{KL}
\end{eqnarray}
where $d\vec{x}$ is the integration measure over the sample space $\{ \vec{x} \}$, and $\Theta_p(\vec{x})$ is the distribution function associated to the values of the parameters
$(\theta^p_1,\ldots,\theta^p_d)$. Now we have defined everything needed to compute the posterior probabilities, and we illustrate
the formalism by applying it to the analysis of light-curves.

Using this, we can compute the Fisher information matrix by computing the
Kullback--Leibler distance between two nearby points and Taylor expanding:
\begin{eqnarray}
D(\Theta_0||\Theta_q) & = & \int d^N\vec{y}\, \,  \Theta_0(\vec{y}) \, \ln \frac{\Theta_0(\vec{y})}{\Theta_q(\vec{y})} \nonumber \\
& \approx & -\int d^N\vec{y}\, \, \Theta_0(\vec{y}) \ln \frac{\Theta_0(\vec{y}) + \partial_{\theta_i} \Theta_0(\vec{y})
\Delta\theta_i + \partial_{\theta_i} \partial_{\theta_j} \Theta_0(\vec{y}) \Delta\theta_i \Delta \theta_j}{\Theta_0(\vec{y})}
\nonumber \\ & \approx & - \int d^N\vec{y} \left( \partial_{\theta_i} \Theta_0(\vec{y}) \Delta \theta_i +
\partial_{\theta_i} \partial_{\theta_j} \Theta_0(\vec{y}) \Delta\theta_i \Delta\theta_j - \frac{1}{2}
\frac{(\partial_{\theta_i}
\Theta_0(\vec{y}))(\partial_{\theta_j} \Theta_0(\vec{y}))}{\Theta_0(\vec{y})} \Delta\theta_i \Delta\theta_j \right) \nonumber \\
& = & \frac{1}{2} \underbrace{\int d^N\vec{y} \frac{(\partial_{\theta_i} \Theta_0(\vec{y}))(\partial_{\theta_j}
\Theta_0(\vec{y}))}{\Theta_0(\vec{y})}}_{\equiv J_{ij}(\Theta_0)} \Delta\theta_i \Delta\theta_j. \Label{eq:fishertheor}
\end{eqnarray}
On the third line, the terms linear in $\Theta_0$ vanish, as exchanging the order of integration and derivation, the integral of
$\Theta_0$ will yield a constant 1, which then differentiates to zero.

\subsection{Effect of the experimental set-up}

The measure (\ref{measure}) is independent of the experimental data $Y$ and is constructed under the assumption that the entire sample space can be measured by the observer.   However, in real experiments, instrumental and design limitations only allow observation of some subset $M$ of the sample space.   Thus an observation either results in no detected outcome, or in a measurement $y_i \in M$.   Thus the effective predicted distribution of measured outcomes is not the $\Theta(\vec{y})$, but rather
\begin{equation}
\Theta(\vec{y}) = \left\{  \begin{array}{ll} \Theta(\vec{y}), & \textrm{for } \vec{y} \in M,
\Label{effdist}
\\ \Theta^{\textrm{Out}} , &
{\rm no \ measured \ outcome},
\end{array} \right.
\end{equation}
where $\Theta^{\textrm{Out}} \equiv \int_{\vec{y} \notin M} d\vec{y} \, \, \Theta(\vec{y})$.     We will argue that if the models in the asymptotic regions of a noncompact parameter space differ in their predictions mostly outside the observable region $M$, the Fisher information for the effective distributions (\ref{effdist}) can decay sufficiently quickly to render the volume $ V = \int d^d\Theta \, \sqrt{\det J_{ij}(\Theta)}$ finite.  In this section we will give one set of sufficient conditions for this to happen and in Sec.~3 we will give a detailed example.

Consider a  model, specified by parameters $\vec{\theta} = (\theta_1, \ldots, \theta_d) \in \mathbb{R}^d$, and a distribution
$\Theta_{\vec{\theta}}(\vec{x})$, with $\vec{y} \in \mathbb{R}^n$. We will slightly simplify notation simply referring to the
distribution as $\Theta(\vec{y})$ and understanding the implicit parameter dependence.  Let us use spherical coordinates in the parameter space $\mathbb{R}^d$ with $\rho$ being the
radial coordinate, i.e. $(\theta_1, \ldots, \theta_d) \to (\rho, \varphi_1,\ldots, \varphi_{d-1})$.  Also consider an experimental set-up that can only make measurements inside some compact region $M \subset \mathbb{R}^n$. Thus,the probability of no measurement being registered by this experiment is $\Theta^{\textrm{Out}} \equiv \int_{\vec{y} \notin
M} d\vec{y}\, \, \Theta(\vec{y})$.

Our first assumption is a smoothness condition, so that inside the region $M$ the distribution does not fluctuate too much as one
approaches the asymptotics of parameter space:
\begin{equation}
\left|\left. \partial_i \Theta(\vec{y})\right|_{\vec{y}\in M} \right| \le \delta(\rho), \quad \textrm{for large } \rho, \, \,
 i = 1,\ldots, d,
\end{equation}
where $\delta(\rho)$ goes to zero as $\rho$ goes to infinity; we will later specify the exact scaling needed.   Intuitively, this
condition says that as the parameter $\rho \to \infty$, the models do not differ too much inside the observable part of the
sample space $M$.  This allows us to estimate
\begin{equation}
\left| \partial_i \Theta^{\textrm{Out}} \right| = \left| \partial_i (1-\int_{y\in M} d\vec{y} \,\, \Theta(\vec{y}))\right| \le
\textrm{Vol}(M) \delta,
\end{equation}
where Vol$(M)$ denotes the volume of the compact region $M$.

Secondly we assume that inside $M$, the distributions $\Theta(\vec{y})$  do not decay too quickly as $\rho \to \infty$.
Intuitively,   since any experiment will only measure a finite amount of data (say $N$ points), if the probability of a single
measurement lying inside $M$ is significantly less than $1/N$, then the experimental set-up will not detect anything. Thus we
will require
\begin{equation}
\Theta(\vec{y})\left.\right|_{\vec{y}\in M} > \epsilon(\rho), \quad \textrm{for large } \rho,
\end{equation}
where again we will later specify the scaling of $\epsilon(\rho)$ with $\rho$.\footnote{This condition can be relaxed by
recognizing that if $\Theta(\vec{y}) |_{\vec{y} \in M}$ decays too quickly as $\rho \to \infty$, then the models in the
asymptotic region of the parameter space make no measurable predictions for experiments designed with a finite number of
measurements.  The example in the Sec.~3 will illustrate such a scenario.}

Using these assumptions, we can establish an upper  bound for the Fisher information (\ref{eq:fishertheor}):
\begin{eqnarray}
\left| J_{ij} \right| & \le & \left| \int_{\vec{y} \in M} \frac{\partial_i \Theta(\vec{y}) \partial_j \Theta(\vec{y})
}{\Theta(\vec{y})}\right| + \left|  \frac{\partial_i \Theta^{\textrm{Out}} \partial_j \Theta^{\textrm{Out}}
}{\Theta^{\textrm{Out}}}\right| \nonumber \\
& < & \delta^2 \left| \int_{\vec{y} \in M} \frac{1}{\Theta(\vec{y})} \right| + \textrm{Vol}(M)^2 \delta^2 \le \textrm{Vol}(M)
\frac{\delta^2}{\epsilon} + \textrm{Vol}(M)^2 \delta^2 \sim \textrm{Vol}(M)\frac{\delta^2}{\epsilon}.
\end{eqnarray}
Thus the determinant of the Fisher information scales as
\begin{equation}
\sqrt{\textrm{Det} \, \, J_{ij}} \sim \left( \frac{\delta^2}{\epsilon} \right)^{\frac{d}{2}},
\end{equation}
and for the integral $V$ to be finite one must have suppression stronger than  $\sqrt{\textrm{Det} \, \, J_{ij}} \sim \rho^{-d}$.
Thus the integral converges if $\delta$ is suppressed more strongly than
\begin{equation}
\delta(\rho) <  \frac{\sqrt{\epsilon(\rho)}}{\rho}.
\end{equation}
From the experimental set-up one can estimate how $\epsilon(\rho)$ scales with $\rho$, which then determines how $\delta(\rho)$
needs to scale for the integral to converge. This is thus a sufficient condition for rendering the
parameter space effectively finite.

It is  worth stressing that, following the above analysis, any method of deciding the validity of a model is impacted by the
choice of the experiment in a completely computable way, and this should be taken into account when designing experiments.

\section{The probability of exo-planet detection}
\llabel{sec:example}

\subsection{Model for exo-planets}

\begin{figure}[t]
\begin{center}
\includegraphics[scale=0.4]{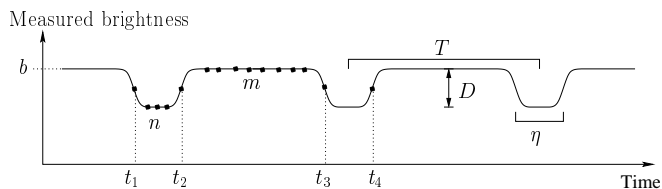}
\caption{An example of a light-curve. }\label{fig:curve}
\end{center}
\end{figure}
Consider a star orbited by a planet so that the planet periodically passes between the star and Earth. The light output
(light-curve) of such a star is a constant line, with a small periodic dip when the planet is eclipsing part of the star.
One model for such a light-curve was proposed in \cite{raul} as
\begin{equation}
y(T,D,\eta,\tau,b;t) = b - \frac{D}{2} \left[\tanh c(\tilde{t}+\frac{1}{2}) - \tanh c(\tilde{t}-\frac{1}{2}) \right],
\Label{eq:curve}
\end{equation}
where
\begin{equation}
\tilde{t} = \frac{T \sin \frac{\pi(t-\tau)}{T}}{\pi \eta}. \Label{t-tilde}
\end{equation}
An example light-curve is shown in figure \ref{fig:curve}; $T$ is the period of the planet; $\eta$ is the duration of the
transit, i.e. how long the planet eclipses the star; $D$ is the depth of the dip in the curve; $b$ is the total observed
brightness of the star; and $\tau$ is a phase parameter specifying when the planets transit occurs. Finally, $c$ is a constant
parameter specifying the sharpness of the edges of the light-curve, expected to be fairly large as the transition between
transit/no-transit is relatively quick. The assumption $c \gg 1$ greatly simplifies our analysis, and is  not physically very
restrictive.

\begin{figure}[t]
\begin{center}
\includegraphics[scale=0.4]{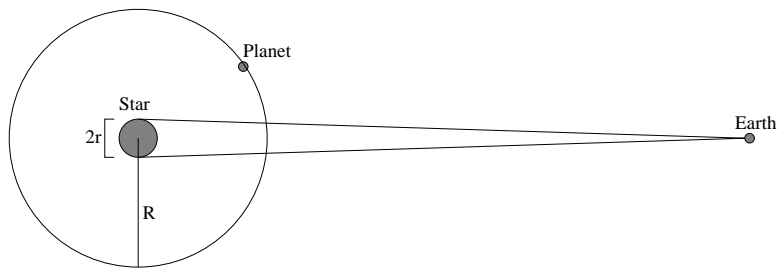}
\caption{The basic set-up: an extra-solar planet orbiting a star of radius $r$ with an average distance $R$. }\label{fig:exo}
\end{center}
\end{figure}

The parameter space for this model is clearly non-compact as $T$ can range to infinity. However, we will argue that the space is
effectively rendered compact after the experimental set-up is specified. To be precise, the parameter space is\footnote{Note that
we consider $c$ to be a constant, not a parameter.}:
\begin{equation}
T \in [0,\infty), \quad D \in [0,b], \quad \tau \in [0,T], \quad \eta \in [0,\delta T], \quad b \in [0,b_{max}],
\end{equation}
where $\delta$ is a small number that we will estimate, and the maximal brightness $b_{max}$ is naturally given by the brightness
of Sirius, the brightest star visible from Earth.  Assuming a circular orbit as in Figure \ref{fig:exo}, the ratio of the transit
time to the period of the planet is given by
\begin{displaymath}
\frac{\eta}{T} \approx \frac{2r/v_{\textrm{planet}}}{ 2\pi R/v_{\textrm{planet}}}  = \frac{1}{\pi} \frac{r}{R}.
\end{displaymath}
For the currently known transiting exo-planets this ratio is around $\sim 0.1$ \cite{data}, although for a typical system one
expects it to be smaller as large planets orbiting close to the star are easier to observe, which favors largest values of the
ratio. For an elliptical orbit, the answer will differ by an $\mathcal{O}(1)$ factor, but will have the same dependence on
$r/R$. Thus, $\eta$ will always be a small fraction of $T$.

Now we can write down the probability density for measuring values $\vec{y} = (y_1, \ldots, y_N)$ for the light-curve at times
$(t_1, \ldots, t_N)$ with the light-curve specified by parameters \newline
\mbox{$(\theta^0_1,\theta^0_2,\theta^0_3,\theta^0_4,\theta^0_5)=(T,D,\eta,\tau,b)$} as
\begin{equation}
\Theta_0(\vec{y}) = \prod_{k=1}^N \frac{1}{\sqrt{2\pi \sigma_k}} e^{-\frac{(y_k - y_0(\theta^0_i;t_k))^2}{2\sigma^2_k}} = (2\pi
\sigma)^{-\frac{N}{2}} e^{-\frac{1}{2\sigma^2}\sum_{k=1}^N (y_k-y_0(\theta^0_i;t_k))^2}, \Label{gauss}
\end{equation}
where we have assumed that the uncertainty in each measurement is Gaussian, and further we have chosen the standard deviation to
be equal for all measurements for simplicity.  Using (\ref{gauss}) in the formula (\ref{eq:fishertheor}), we see that the   the
integrals in the Fisher information  are Gaussian in $y_k$; thus we can compute them analytically to get
\begin{equation}
J_{ij} = \frac{1}{\sigma^2} \sum_{k=1}^N \partial_{\theta_i} y(\theta; t_k) \partial_{\theta_j} y(\theta;t_k). \Label{eq:fisher}
\end{equation}
This is our key formula, and we shall spend the next subsection analysing its properties.

\subsection{Finiteness of light-curve parameter space}
\llabel{finiteness} We now wish to apply the general arguments of section \ref{general-argument} to the exo-planet system.
Consider an experimental set-up that can barely measure two periods, and then consider shortening the experiment slightly so that
only one dip is detected; this is depicted in figure \ref{fig:curve}. To be precise, the shorter set-up measures the beginning
and end of a transit at $t_1$ and $t_2$, $n$ points in between, and $m$ points after the transit. The longer set-up makes
measurements at the same times, and additionally at times $t_3$ and $t_4$, detecting the second transit. In the next subsections
we will show that $J_{\textrm{short}} \lll J_{\textrm{long}}$, indicating that detecting the second dip is of fundamental
importance to experimental design; without the second dip the experimental set-up can't differentiate models with large enough
$T$. This renders the parameter space effectively finite, as an experiment can not differentiate between models that have period
$T$ larger than the duration of the experiment.

\subsubsection{Effect of measuring a second transit on det$(J)$}
\llabel{sec:estimate} In this subsection we will give an estimate for the magnitude of the determinant of the Fisher information,
and show how it is affected by the inclusion of the second transit in the data. In subsequent subsections we will exactly compute
the determinant for a few specific experimental set-ups.

From (\ref{eq:fisher}) and the definition of a determinant, we see that in each term of the determinant each parameter $\theta_i$
appears exactly twice in the derivatives, i.e. each term is of the form
\begin{equation}
J_{i_1j_1}J_{i_2j_2}J_{i_3j_3}J_{i_4j_4}J_{i_5j_5} \sim \frac{1}{\sigma^{10}}\partial_T y\, \,  \partial_T y\, \, \partial_D y\,
\,
\partial_D y\, \,  \partial_{\eta} y\, \,  \partial_{\eta} y\, \,  \partial_{\tau} y\, \,
\partial_{\tau} y\, \, \partial_b y \, \, \partial_b y. \Label{generic}
\end{equation}
As a rough estimate of the size the determinant, we investigate how large terms of this type can be. The derivatives are
\begin{eqnarray}
\partial_T y(\theta,t) & = & \frac{cD}{2}  \frac{f(\tilde{t})}{\pi\eta}\left(\sin \frac{\pi(t-\tau)}{T} - \frac{\pi(t-\tau)}{T} \cos
\frac{\pi(t-\tau)}{T}\right),  \Label{DT} \\
\partial_{D} y(\theta,t) & = & \frac{1}{2} \left(\tanh c(\tilde{t}-\frac{1}{2}) - \tanh c(\tilde{t}+\frac{1}{2}) \right),
\Label{DD} \\
\partial_{\tau} y(\theta,t) & = & -\frac{cD}{2} \frac{f(\tilde{t})}{\eta} \cos \frac{\pi(t-\tau)}{T}, \Label{Dtau} \\
\partial_{\eta} y(\theta,t) & = & -\frac{cD}{2} \frac{\tilde{t} f(\tilde{t})}{\eta}, \quad \quad \partial_b y(\theta,t) = 1, \Label{Deta} \\
\textrm{with} & & f(\tilde{t}) \equiv \tanh^2 c(\tilde{t}+\frac{1}{2}) -  \tanh^2 c(\tilde{t}-\frac{1}{2}). \Label{eq:f}
\end{eqnarray}
From (\ref{eq:f}) we see that $f(\tilde{t}) \neq 0$ only when $\tilde{t} \approx \pm \frac{1}{2}$, again assuming large $c$. This
tells us that the measurements that contribute most to the Fisher information are the ones on the edges of the dips\footnote{This
statement is somewhat subtle, and we will discuss this matter in more detail in section \ref{sec:dets}; for our current purposes
it is sufficiently accurate.} , i.e. at times $t_1,t_2,t_3$ and $t_4$ in figure \ref{fig:curve}. We write the condition
$|\tilde{t}| \approx \frac{1}{2}$ as
\begin{equation}
\left| \sin \frac{\pi (t-\tau) }{T} \right| = \frac{\pi}{2} \frac{\eta}{T},
\end{equation}
and note that the ratio of transit time to period is very small, $\frac{\eta}{T} \ll 1$. This gives us the solutions
\begin{equation}
\frac{t-\tau}{T} \approx n \pm \frac{\eta}{2T},  \Label{valley-index}
\end{equation}
where $n$ is an integer indexing the number of the dip, with $n=0$ denoting the solitary dip if only one is present in the data.

We wish to estimate the ratio of the determinants of the Fisher information by an order of magnitude estimate
\begin{equation}
\frac{J_{\textrm{short}}}{J_{\textrm{long}}} \sim \frac{J_{i_1j_1}^sJ_{i_2j_2}^sJ_{i_3j_3}^sJ_{i_4j_4}^sJ_{i_5j_5}^s \left.
\right|_{\textrm{max}}}{J_{i_1j_1}^lJ_{i_2j_2}^lJ_{i_3j_3}^lJ_{i_4j_4}^lJ_{i_5j_5}^l \left. \right|_{\textrm{max}}},
\Label{ratio}
\end{equation}
where both the numerator and the denominator are of the form (\ref{generic}), and according to the argument above the maximal
contributions come from the edge measurements. From  (\ref{DD}-\ref{Deta}) we see that the derivatives with respect to $D,\eta,
\tau$ and $b$ are all periodic at the edges: $|\partial_{\theta_i} y(\theta,t_1)| = \ldots = |\partial_{\theta_i} y(\theta,t_4)|$
for $\theta_i \neq T$, and thus will cancel in the ratio (\ref{ratio}).

It is crucial that $\partial_T y$, however, is not periodic due to the second term in (\ref{DT}). At the first dip, $t_1, t_2 =
\pm \frac{\eta}{2T}$, we expand (\ref{DT}) to find
\begin{equation}
\partial_T y(t_1) \approx \partial_T y(t_2) \approx \frac{\pi^2 cD}{48} \frac{\eta^2}{T^3},
\end{equation}
while at the second dip, $t_3,t_4 = 1 \pm \frac{\eta}{2T}$, the contribution is
\begin{equation}
\left| \partial_T y(t_3) \right| \approx \left| \partial_T y(t_4) \right| \approx \frac{cD}{2\eta},
\end{equation}
ignoring signs that are irrelevant for this estimate. Thus we see that the Fisher information increases strongly as the second
dip is included:
\begin{equation}
\frac{J_{\textrm{short}}}{J_{\textrm{long}}} \sim \frac{(\partial_T y(t_{1,2}))^2}{(\partial_T y(t_{1,2}))^2 + (\partial_T
y(t_{1,2}) \partial_T y(t_{3,4}) ) + (\partial_T y(t_{3,4}))^2} \sim \left( \frac{\eta}{T} \right)^6 \lll 1,
\end{equation}
where we ignored order one coefficients. This is an explicit example of how our arguments from section \ref{general-argument}
work for a realistic model: when an experimental set-up does not have the capability to detect two dips, it becomes impossible to
determine the period, and consequently the Fisher information is very small (or vanishing) compared to an experiment that is able
to detect two dips and determine the period more accurately.   For any given experiment of finite duration $\Delta t$, the Fisher Information will decline with $T$ when $T \gg \Delta t$ effectively rendering the parameter space compact.

\subsubsection{The tail $T \to \infty$}
To verify our claim that the parameter space is really rendered compact we need to show that det $J \to 0$ strongly enough as $T$
is taken to infinity. It is easy enough to find the $T$-scaling of the derivatives  (\ref{DT}-\ref{Deta}); $\partial_T y$ scales
as $T^{-3}$, while the others stay finite in the large $T$ limit. Thus, as seen from (\ref{generic}), the determinant will scale
as
\begin{equation}
\sqrt{\textrm{det } J} \sim \partial_T y \sim \frac{1}{T^3},
\end{equation}
which shows that that the parameter space measure vanishes fast enough for large $T$ to render the parameter space volume finite.

\subsubsection{Explicit computation of Det$(J_{ij})$ for specific experimental set-ups}
\llabel{sec:dets} While the order of magnitude estimate of the previous subsection offers an intuitive reason as to why the
Fisher information decreases sharply when the number of peaks detected falls below two, it is still instructive to explicitly
compute the determinant in a few experimental set-ups.

\paragraph{Detecting two dips:} Let us first consider the case $J_{\textrm{long}}$ from section \ref{finiteness}, i.e. measurements at times indicated in
figure \ref{fig:curve}.  Using the derivatives (\ref{DT}-\ref{Deta}) one can write down the Fisher information matrix
(\ref{eq:fisher}) as
\begin{equation}
J^{\textrm{long}}_{ij} = \left( \begin{array}{ccccc} 2(T_1^2+T_3^2) & -T_1 & 2T_1X & -4T_3X & 2T_1 \\ -T_1 & 1+n & -2X & 0 &
-(2+n) \\ 2T_1X & -2X & 4X^2 & 0 & 4X \\ -4T_3X & 0 & 0 & 16X^2 & 0 \\ 2T_1 & -(2+n) & 4X & 0 & 4+n+m
\end{array} \right),
\end{equation}
where for brevity we defined
\begin{eqnarray}
T_1 & \equiv & \partial_Ty(t_1) = \partial_Ty(t_2) = \frac{cD\pi^2}{48}\frac{\eta^2}{T^3} , \quad T_3 \equiv \partial_Ty(t_3) =
-\partial_Ty(t_4)
= \frac{cD}{2\eta}, \\
X & \equiv & -\frac{cD}{4\eta} = \partial_{\eta}y(t_{1,2,3,4}) = -\frac{\partial_{\tau}y(t_{1,3})}{2} =
\frac{\partial_{\tau}y(t_{2,4})}{2}.
\end{eqnarray}
In computing this matrix we used that $f(\tilde{t})=0$ for $\tilde{t}\neq \pm \frac{1}{2}$, which is true up to corrections of
order $e^{-c}$, as seen from (\ref{eq:f}); for this reason one does not need to specify the exact times of the $n$ measurements
during the dip, or the $m$ measurements outside the dip, as up to $e^{-c}$ corrections they all contribute equally. The
determinant of the Fisher information is simple,
\begin{equation}
\textrm{Det} (J_{ij}^{\textrm{long}}) = 64 nm X^4 (T_1^2+T_3^2) \approx 64 nm X^4 T_3^2.
\end{equation}
This result explains the subtlety referred to earlier: although measurements at the edges contribute the most to the Fisher
information, if one only has measurements at the edges ($n=m=0$) the Fisher information actually vanishes. Physically this is
easy to interpret, as only measuring the edges $t_1,\ldots,t_4$ will yield four points lying on a line, and thus they cannot be
used to determine any information about the curve; other data points are needed to `anchor' the data.

\paragraph{Detecting only one dip:} Similarly one can compute the Fisher information in the `short' experimental set-up, where
measurements are made at the same times as before, except not at $t_3$ and $t_4$. This yields
\begin{equation}
J^{\textrm{short}}_{ij} = \left( \begin{array}{ccccc} 2T_1^2 & -T_1 & 2T_1X & 0 & 2T_1 \\ -T_1 & \frac{1}{2}+n & -X & 0 & -(1+n)
\\ 2T_1X & -X & 2X^2 & 0 & 2X \\ 0 & 0 & 0 & 8X^2 & 0 \\ 2T_1 & -(1+n) & 2X & 0 & 2+n+m
\end{array} \right),
\end{equation}
and perhaps surprisingly the determinant vanishes: Det$(J^{\textrm{short}}_{ij}) = 0$, up to tiny $e^{-c}$ corrections.  This
indicates that the estimate in section \ref{sec:estimate} was an overestimate\footnote{As the estimate illustrates an intuitive
reason why the appearance of the second peak is so important, we decided to include it.}: terms in the determinant of
$J^{\textrm{short}}$ are of the magnitude estimated, but the determinant is arranged in such a way that the terms cancel to a
high accuracy, and the compactness of the parameter space is strengthened.

\section{Discussion}

Our analysis has shown how the specification of an experimental design affects the measure on model parameter spaces in MDL model
selection  (or equivalently the prior probability distribution on parameters in the Bayesian approach).   Interestingly, the
finite number of measurements within a bounded sample space in any practical experiment can effectively render a non-compact parameter space compact thereby
leading to a well-defined prior distribution (\ref{measure}).   Our analysis could be turned around to design experiments to
discriminate well between models in some chosen region of the parameter space by ensuring that the Fisher information
(\ref{eq:fisher}) is large in the desired region.  It would also be useful to determine general conditions under which experimental design effectively makes model parameter spaces compact, perhaps following the arguments of Sec.~2.

\paragraph{Acknowledgments: } This paper was written in honor of Jorma Rissanen's 75th birthday and his many seminal achievements in
statistics and information theory.  VB and KL were partially supported by the DOE under grant DE-FG02-95ER40893, and KL was also
partly supported by a fellowship from the Academy of Finland. VB was also partly supported as the Helen and Martin Chooljian
member at the Institute for Advanced Study.

\end{document}